\documentclass[11pt,a4paper]{article}
\pdfoutput=1
\usepackage{jcappub}
\usepackage{graphicx}
\usepackage{subfig}

\def\apj{{\it Astrophys.~J.}}
\def\apjs{{\it Astrophys.~J. Suppl.}}

\def\apjl{{\it Astrophys.~J.~Lett.}}
\def\prd{{\it Phys.~Rev.~D}}
\def\prl{{\it Phys.~Rev.~Lett.}}

\def\mnras{{\it Mon.~Not. Roy.~Astr.~Soc.}}
\def\ijmpd{{\it Int.~J.~ Mod. Phys. D}}
\def\nature{{\it Nature}}

\def\AnA{{\it Astron. Astrophys.}}
\def\grg{{\it Gen. Rel. Grav.}}
\def\ARAnA{{\it Ann. Rev. Astron. Astrophys.}}

\title{Cosmic distance duality and cosmic transparency}
\author[a]{Remya Nair,}
\author[a]{Sanjay Jhingan}
\author[b]{and Deepak Jain}

\affiliation[a]{Centre for Theoretical Physics,\\ Jamia Millia
Islamia, New Delhi 110025, India}

\affiliation[b]{Deen Dayal Upadhyaya College, \\
University of Delhi, New Delhi 110015, India}

\emailAdd{remya$_{-}$phy@yahoo.com}
\emailAdd{sanjay.jhingan@gmail.com} \emailAdd{djain@ddu.du.ac.in}

\abstract{We compare distance measurements obtained from two distance indicators, Super-
novae observations (standard candles) and Baryon acoustic oscillation data (standard rulers).
The Union2 sample of supernovae with BAO data from SDSS, 6dFGS and the latest BOSS
and WiggleZ surveys is used in search for deviations from the distance duality relation. We
find that the supernovae are brighter than expected from BAO measurements. The luminos-
ity distances tend to be smaller then expected from angular diameter distance estimates as
also found in earlier works on distance duality, but the trend is not statistically significant.
This further constrains the cosmic transparency.}

\keywords{distance duality, cosmic acceleration, supernovae, baryon
acoustic oscillations, cosmic opacity }

\begin{document}

\maketitle

\section{Introduction}
The last few decades have been truly exciting in the field of
Cosmology. With the success of many cosmological surveys and many
more lined up in near future, we are in an era of data driven
Cosmology \cite{sur}. Most of these observations hint at a Universe
going through a phase of late time accelerated expansion \cite{DHwein}. There have
been many attempts to understand the physical interpretations of
these measurements \cite{exp,ruth}, within and beyond the framework
of general relativity (GR). In the framework of GR the component
that drives the cosmic expansion is termed as {\em Dark Energy}. The
simplest and most successful candidate for dark energy  is the
cosmological constant ($\Lambda$), whose energy density remains
constant with the expanding Universe. However, $\Lambda$ suffers
from serious problems like fine tuning and the coincidence problem
\cite{wein}. Alternatively, a evolving dark energy scenario can be
explained by using scalar field models (See \cite{sami} for a
review). Alongside there have been attempts to explain the observed
accelerated expansion by modifying gravity at large scales (see
\cite{kunz} for a recent review and references therein).

Cosmological studies are based on some fundamental assumptions like
the Copernican principle, conservation of photons etc.
There is no evidence of structure formation at scales larger than
100 Megaparsec and hence large scale homogeneity of the Universe is
a common assumption in Cosmology. The almost uniform Cosmic
Microwave Background (CMB) hints at a Universe that is isotropic as
well. Building a robust cosmological model requires us to validate
these assumptions. There have been many attempts to constrain the inhomogeneity
of the Universe. Using observations from Sunyaev-Zeldovich effect in
clusters of galaxies, excitation of low-energy atomic transitions,
and the thermal spectrum of the CMB, Goodman \cite{good} examined
the radial homogeneity of the Universe. Caldwell et al., \cite{cald}
used the blackbody nature of the CMB spectrum to test the Copernican
principle and placed a limit on the possibility that we occupy a
privileged location. Clarkson et al., \cite{clark} proposed an
observational test for the Copernican assumption based on a
consistency relation of the Friedmann-Lemaitre-Robertson-Walker
(FLRW) model between cosmic distances and the Hubble expansion rate.
Further, there have been attempts to constrain the transparency of
the Universe at visible and cm wavelengths \cite{sri,sandage}. More
et al. \cite{mor}, used the Supernovae data and the Baryon acoustic feature to
constrain the optical depth of the universe, and
Avgoustidis et al. \cite{avg1}, constrained the
cosmic transparency by combining Supernovae with Hubble rate data.

The dynamical properties of the Universe can be deduced by analysing
multiple cosmological observations. Moreover, various independent
measurements can be combined to find better constraints on our
estimates of the Cosmological parameters. It is important to device
consistency probes using these measurements to verify the underlying model and
its assumptions. In this work we
analyse a consistency relation between distances: the Distance
Duality relation (DD). It relates the angular diameter distance
(ADD) to the Luminosity distance (LD). DD is a model independent
probe that can be used to compare and chose between competing
cosmological models. Bassett and Kunz \cite{bk} used it to rule out
the replenishing grey-dust model which causes redshift-dependent
dimming of the Supernovae (no cosmic acceleration). Additionally,
deviation from DD, if observed, would hint at a breakdown of one or
more of its underlying assumptions. One of the assumptions is photon
number conservation. There could be both clustered and un-clustered
sources of photon attenuation in the Universe. Clustered sources
like gas and plasma in the galaxies are correlated with large scale
structures. More exotic sources causing change in photon number are axions or
axion like particles, motivated from string theory models.
Photon-axion coupling can give rise to Supernovae dimming, but such
scenario are strongly constrained by cosmological data \cite{axion}.
Unclustered sources of attenuation effect all the lines of sight
equally and can be detected by comparing radiation sources of known
properties at different redshifts. We build here on the method
proposed by More et.al \cite{mor} to put constraints on the cosmic
transparency using the current Baryon Acoustic Oscillations (BAO)
and the SNeIa data.

The plan of the paper is as follows. In section 2 we give a quick
overview of DD. This is followed by section 3 on cosmic
transparency. Section 4 summarises data sets used, with subsections
on constraints on distance duality and cosmic transparency. We conclude
with a section discussing main results.

\section{The distance duality relation}

In 1933 Etherington showed that a relation exists
for area distances
between a source and an observer in relative motion with each other.
He proved a reciprocity relation valid for any curved spacetime,
i.e., it does not assume even the basic symmetry assumptions like
homogeneity and isotropy \cite{eth}. The
relation holds as long as gravity is described by a metric theory,
photons travel on null geodesics and the geodesic deviation equation
is valid \cite{bk,uz}. Further, if photon number is conserved one
can relate  the angular diameter distance (ADD)
and the Luminosity distance (LD) using the reciprocity relation
\cite{dd},
\begin{equation}
d_{L}=d_{A} (1+z)^{2}.
\end{equation}
This is termed as the Distance-Duality (DD) relation and plays an
important role in galaxy cluster observations and lensing studies
\cite{ellisgrg}. A consequence of DD is the optical theorem which
states that the surface brightness of an object depends on its
redshift and not on its distance from the observer. This allows us
to derive a temperature shift relation for CMB photons relating their observed
and emitted temperature via $T_{o}=T_{e}/(1+z)$.

Evidently DD is  crucial to Cosmological studies and plays a
key role in how galaxy observations are analysed. Hence, it is
important to check its validity. This is possible in principle since
both the distances in the duality relation are observable. Defining
\begin{equation}
\eta (z)\equiv \frac{d_{L}}{d_{A}(1+z)^{2}},
\end{equation}
the DD holds if $\eta (z)$ = 1 at all redshifts. As stated earlier
the violations in DD could arise due to change in photon number,
if there are deviations from metric
theory of gravity, or if photons do not travel on unique null
geodesics. For our analysis here we look for violations in
DD due to a less extreme possibility: violation of photon number
conservation.

There are various ways of measuring cosmic distances. One has
to look for sources that can be used as standard candles for
deriving LD, and as standard rulers for deriving ADD. Type Ia
Supernovae (SNeIa) have a peak luminosity that is tightly correlated
with the shape of their light curves and they can be calibrated to
be treated as standard candles. Combined measurements of the
Sunyaev-Zeldovich effect and X-Ray analysis (SZE/X-ray) provides a
measure of the ADD to a cluster. Other sources that can be used for
ADD estimates are FRIIb radio galaxies, compact radio sources etc.
The Baryon acoustic feature in the matter clustering is another
independent distance indicator and can be used as a standard ruler.

Bassett $\&$ Kunz \cite{bk} were the first to propose DD as a
powerful tool for  testing  exotic physics. They used SNeIa data for LD, and
FRIIb radio galaxies, compact radio sources and X-ray clusters for
ADD to verify the duality relation. Uzan et al., \cite{uz} showed
that the X-Ray/SZ combined analysis of galaxy clusters offers a
validity test of DD. More importantly they proved that the analysis
does not give a measurement of the ADD if DD is violated. Since then
there have been many attempts to constrain violations in DD. De
Bernardis et al. \cite{b} used Chandra cluster data and found no
evidence for DD violation within the framework of $\Lambda$CDM.
Corasaniti \cite{co} modeled the intergalactic dust in terms of the
star formation history of the universe and forecasted a deviation in
the DD relation due to the presence of the cosmic dust extinction.
Lazkoz et al., \cite{la} used  SNeIa and CMB + BAO as standard
candles and standard rulers respectively to test the validity of
this relation and found $\eta$ = 0.95 $\pm$ 0.025 in the redshift
range $0 < z < 2$, which is consistent with distance duality at the
2$\sigma$ level. Holanda et al.\cite{h0} checked the validity of DD
relation using two different galaxy cluster samples and SNeIa data
and constrained violation in DD by parameterizing $\eta$(z). They
concluded that the best fit values of the parameters of the
$\eta$(z) parametrization obtained through the data set based on
spherical $\beta$ model of cluster are not consistent with the DD
relation. We found similar results in our earlier work on DD using
different distance measures for the ADD \cite{us}. There have been
many other attempts to verify/constrain DD or use it to put
constraints on different cosmological parameters  \cite{restdd}.

The standard practice to check the consistency of DD is to use
the LD estimate from the SNeIa data, and get corresponding
ADD estimate from one of the aforementioned sources. There are two main
sources of error in such a analysis. Firstly, it is not
possible to find SNeIa and galaxy clusters at exactly the same
redshift and  some kind of selection criteria (eg. $\Delta z =
|z_{cluster}-z_{SNeIa}| < 0.005$) is applied.  The redshift mismatch
between the cluster and the SNeIa may introduce some errors in the
analysis. Secondly, the measurement of ADD from clusters, as described
above, is effected by the assumption on cluster geometry. It has
been shown earlier that different assumptions regarding cluster
geometry gives significantly different constraints on the deviation
from DD (see for example \cite{h0,us}).
Thus if DD holds it can act as probe of cluster geometry.

To study deviation from DD in a model independent way  we
parameterize $\eta$ as follows:
\begin{align}
\eta(z) &= 1+\eta_{0}z,\\
\eta(z) &= 1+\eta_{0} \frac{z}{1+z}
\end{align}
The first one is a simple Taylor series expansion near low $z$ but
is ill behaved at higher $z$ values. The second is well behaved even
at really high redshifts and is slowly varying as compared to the
first one. For both the parameterizations DD is recovered if $\eta
_{0} = 0$. Note here that though such parameterations are a useful
model independent tool yet a particular form for $\eta(z)$ puts a strong prior.
As we do not expect large variation in the value of $\eta$ over a small redshift
range a piecewise cosntant parametrization can be particularly useful here.
Since we have a small data set of seven BAO observations we divide the redshift
range in three bins with boundaries: $0 <z < 0.35$, $0.35 <z < 0.7$ and $0.7 <z < 1$.

We will analyse one more parameterization (see next section for details)
to account for deviations in DD due to violation in photon number conservation.

\section{Cosmic transparency}

The measurement of the CMBR intensity spectrum is one of the most
precise measurements in Cosmology today \cite{mather,koma}. The
temperature redshift relation mentioned earlier, relating the
observed and emitted temperature of the CMB photons, assumes photon
conservation. The relation will be modified if this assumption was
violated. Many mechanisms have been proposed which may give rise to
such a violation, for example decaying vacuum cosmology, photon
axion coupling etc. (see \cite{avg2} and references therein). There
have been attempts to measure the CMB temperature at different
redshifts, using, for example, quasar absorption line spectra
\cite{sri}. This can be used to check the validity of the
temperature shift relation and hence the validity of the assumption
of photon conservation. But the uncertainties are large and more
data is required to put robust constraints. Here we look at an
alternate method to find evidence for violation of photon
conservation and consequently a violation in DD.

In this section we build on a method proposed by More et al.,
\cite{mor}, to constrain the opacity of the Universe. As mentioned
before if there is some kind of photon-absorption in the Universe
then DD will be violated. If one assumes that there are some
unclustered sources of photon attenuation in the Universe, and
$\tau$(z) is the opacity between the source (for example a
Supernovae) at some redshift $z$ and the observer, then the flux
observed at $z=0$ is reduced by a factor of $e^{-\tau(z)}$ or
equivalently\begin{equation}
d_{L,obs}^{2}=d_{L,true}^{2} e^{\tau (z)}. \label{dl}
\end{equation}
More et al., used the BAO data at redshifts 0.2 and 0.35 along with
Supernovae data to constrain the opacity between these redshifts.
Defining $\Delta\tau$ as the difference in optical depth between the two
redshifts, assuming DD, in a flat-$\Lambda$CDM Universe and with a
prior $0< \Delta \tau <0.5$, they found that $\Delta \tau <0.13$ at
$95\%$ confidence. But the best fit value was negative indicating
that there was slight disagreement between BAO and SNeIa
measurements and that the Supernovae are brighter than expected if
estimated using BAO data (assuming DD).

Avgoustidis et al.,\cite{avg1} proposed a parameterization of $\eta (z)$
to account for photon attenuation
\begin{equation}
\eta(z) = (1+z)^{\epsilon} .
\end{equation}
Here $\epsilon$ is a parameter that characterizes departure from DD. They found
results similar to More et al., using SNeIa and Hubble data.  Since $\epsilon$ can be related to the optical depth parameter $\Delta\tau$ as $\tau
= 2 \epsilon z$ for low $z$, they gave an upper bound on the opacity  $\Delta \tau < 0.02$. We also analyse this parametrization to put constraints on $\epsilon$. Further, we follow the strategy used in \cite{mor} and constrain the cosmic opacity between various redshifts using the BAO and Supernovae data.

If distance duality holds then $\eta$ is a constant in time
(redshift), i.e. $\eta (z_{1})=\eta (z_{2})$, implying
\begin{equation}
\frac{d_{L}(z_{1})}{d_{A}(z_{1}) (1+z_{1})^{2}} =
\frac{d_{L}(z_{2})}{d_{A}(z_{2}) (1+z_{2})^{2}} \;.
\label{one}
\end{equation}
The distance modulus and luminosity distance are related ($d_L(parsec)=10^{0.2 \mu +1}$), and most galaxy cluster surveys
provide measurements of an angle-averaged distance $D_{V}$
\begin{equation}
D_{V}=\left(\frac{cz(1+z)^{2}d_{A}^{2}}{H(z)}\right)^{\frac{1}{3}},
\label{da}
\end{equation}
or the distilled parameter $d_z=r_{s}/D_{V}$ \cite{bl1,bl2}.
Using (\ref{dl}) and (\ref{da}) in (\ref{one}) we can derive the
difference in opacity between two redshifts $z_1$  $\&$ $z_2$ as given in (\cite{mor}),
\begin{equation}
\Delta \tau = \frac{\ln(10)}{2.5} \left[\Delta \mu _{obs} + 2.5 \log
\left(\frac{z_{2}(1+z_{1})^{2}H(z_{1})}{z_{1}(1+z_{2})^{2}H(z_{2})}
\left(\frac{d_{z}(z_{2})}{d_{z}(z_{1})}\right)^3 \right) \right] \;.
\end{equation}
which is written in terms of the distilled parameter $d_z$.
Here $\mu (z)$ is the distance modulus corresponding to redshift
$z$ as obtained from the Supernovae data and $\Delta \mu _{obs} =
\mu (z_{2})- \mu (z_{1})$.

\section{Constraints from current data}

In this work we use the BAO measurements as a standard ruler for
inferring ADD \cite{eis,coo}. This data is more reliable since the
BAO physics is well understood and is not effected by the
systematics mentioned earlier. To obtain $\eta (z)$ we require
the observed values of LD and ADD at the same redshift. This can
be achieved by redshift-matching, as mentioned earlier. Further,
a solution to the redshift-mismatch problem can be found using local regression technique by
inferring the distance modulus of the SNeIa at the same redshift as the
effective redshift of the BAO measurement \cite{car}.

\subsection{Data}
For estimating LD we use the Union2 sample of type Ia
Supernovae \cite{aman}, and BAO data from different galaxy cluster
surveys - SDSS ($z$=0.2, 0.35), 6dFGS($z$=0.106), WiggleZ($z$=0.44, 0.6, 0.73) and BOSS($z$=0.57) for estimating ADD  \cite{bl1,perc,beut,boss1}.
BAO in the observed galaxy power spectrum has a characteristic
scale ($r_s(z_{*})$) determined by  the comoving sound horizon at an
epoch ($z_*$) slightly after decoupling. This epoch is measured by
CMB anisotropy data \cite{koma}. The strength of BAO over other
probes lies in the fact that it can be measured in both radial
(giving $r_s H(z)$) and transverse directions (giving $d_A/r_s$).
Here $H(z)$ is the Hubble expansion rate, $d_A$ is the ADD. Since
$r_{s}$ is well constrained by the CMB data, it is possible  to get
a direct estimates of $d_{A}$ and $H(z)$ from transverse and radial BAO measurements.

We analyse three data sets. The estimate of distance modulus in all three data sets comes from the Union2 sample. The estimate of the angular diameter distance in data set I and II come from SDSS,6dFGS WiggleZ and BOSS and the difference in the two data sets arises due to the different methods used to find the LD estimate corresponding to the ADD estimate. Data set III uses $d_A$ estimates from some recent papers. Our main analysis will be based on data set II but we show some results obtained from data set I and III for comparison. Now we briefly discuss the three data
sets:

\begin{itemize}
\item Data set I (Redshift matching) : If a Supernova is not available  at a particular BAO redshift,
a nearby Supernova satisfying $\Delta z = |z_{BAO}-z_{SNeIa}| <
0.005$ is chosen. Here $z_{BAO}$ and $z_{SNeIa}$ are the BAO
redshift and redshift of the Supernova, respectively. \\ The estimate on $d_A$ can be derived from $d_z$ within the framework of
$\Lambda$CDM assuming $r_s$ = 153.3$\pm$2.0 Mpc ,  $\Omega_m$ = 0.283 $\pm$ 0.017 and $H_0$ = 69.3
$\pm$ 1.5Km/s/Mpc \cite{boss}. We have 7 data points in this set.

\item Data set II (Local regression) : The distance modulus can be obtained at the BAO redshift by
using a local regression method. Following Cardone et al.,\cite{car} we form subsamples of
the Union2 data by choosing $14$ Supernovae which are nearest to the
BAO redshift. We then fit a first order polynomial to this subsample
weighing each Supernova with its corresponding weight where the
weighing function is

\begin{equation}
W(Z) = \left \{
\begin{array}{ll}
(1 - |Z|^3)^3 & |Z| \le 1 \\ ~ & ~ \\ 0 & |Z| \ge 1 .
\end{array}
\right .
\end{equation}
Here $Z = |z_{cl} - z_i|/\Delta$ and $\Delta$ is the maximum value
of the $|z_{BAO} - z_i|$ in the subsample. The zeroth order term is
chosen as the estimate of the distance modulus at the redshift
$z_{BAO}$. The error on the distance modulus is estimated as the
r.m.s value of the residuals (weighted) with respect to the best
fit. The $d_{A}$ estimates are the same as in data set I. This data set also contains 7 data points.

\item Data set III : In this data set $d_L$ is estimated using local regression as
discussed above. For $d_A$(z) estimates we use the values
given in \cite{xu,bl3,boss}. Reid et al.
\cite{boss} give an estimate of $d_{A}$ for $z=0.57$, Xu et al. \cite{xu}
use the anisotropy of the BAO signal measured in the galaxy
clustering distribution of the Sloan Digital Sky Survey (SDSS) Data
Release 7 (DR7) Luminous Red Galaxies (LRG) sample and apply
density-field reconstruction to an anisotropic analysis of the
acoustic peak to give estimate of $d_A$ at $z=0.35$. Blake et al.
\cite{bl3} give estimates of $d_A$ at redshifts z = 0.44, 0.6 and
0.73 by combining measurements of the acoustic parameter $ A \propto
(D^{2}_{A}/H)^{1/3}$  and Alcock-Paczynski distortion parameter $F
\propto D_{A} H$ from galaxy clustering in the WiggleZ Dark Energy
Survey. This data set has 5 data points.

\end{itemize}

\subsection{Constraints: Cosmic distance duality}

We obtain $\eta ^{obs}=d_{L}/d_{A}(1+z)^2$ using estimates of $d_{L}$ and $d_{A}$
as discussed above for the three data sets. The error on $\eta
^{obs}$ is obtained by combining errors from SNeIa
measurement and BAO measurements (data set I and II have additional
errors from the Cosmological parameters
$r_{s}$, $\Omega_{m}$ and $H_{0}$). We assume that the errors have a
gaussian distribution and find the best fits to the parameters by
minimizing  the chi-squared function:
\begin{equation}
\chi^{2}(p)=(\eta ^{obs}-\eta ^{th}(p))^T C^{-1}(\eta ^{obs}-\eta ^{th}(p)),
\end{equation}
where ($\eta ^{obs} - \eta ^{th}(p)$) is the vector of the
differences between the observed and theoretical  $\eta$. The model
parameters ($\eta _{0}$ or $\epsilon$) are denoted by $p$. $C$ is the covariance matrix evaluated using the
correlation coefficients between different redshift slices in the
BAO data. We present our best fits with 1$\sigma$ error bars in the
table \ref{tab1} below. Figure \ref{figa} summarises the results of this section.

\begin{figure}[ht]
\centering \subfloat[Part 3][]{
\includegraphics[bb=0 0 288
286,width=3in]{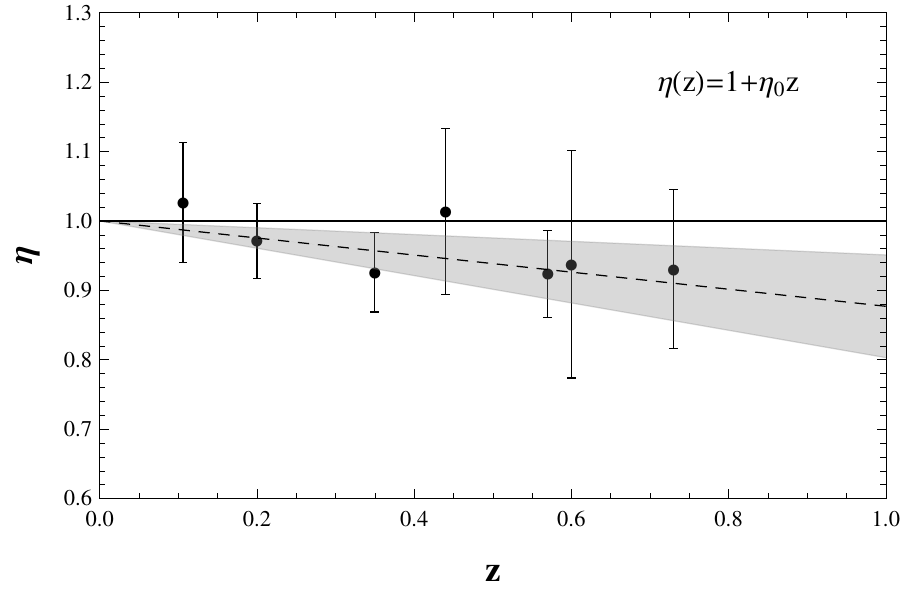}\label{fig11}} \subfloat[Part
3][]{\includegraphics[bb=0 0 288
286,width=3in]{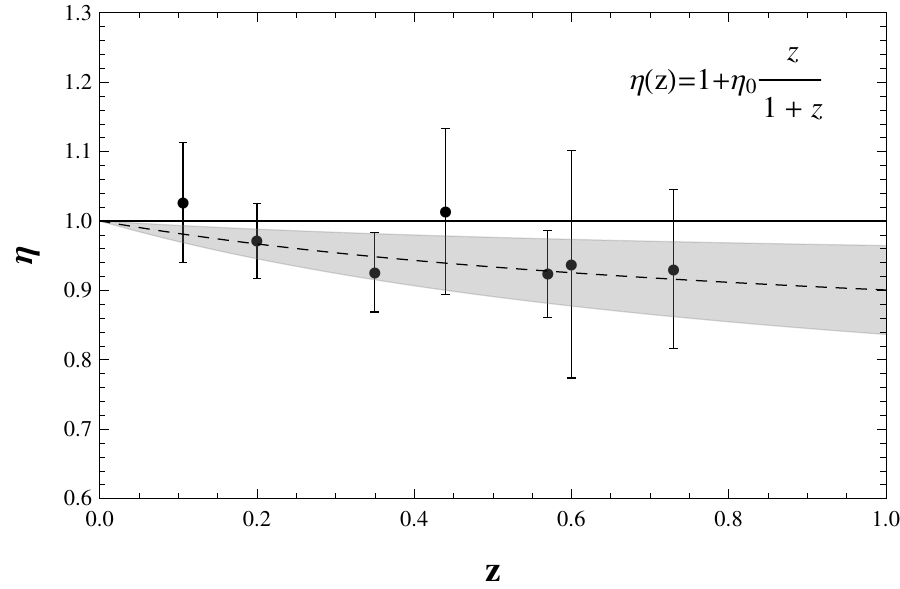}\label{fig12}} \\ \vspace{-0.8in}\subfloat[Part
3][] {\includegraphics[bb=0 0 288
286,width=2.5in]{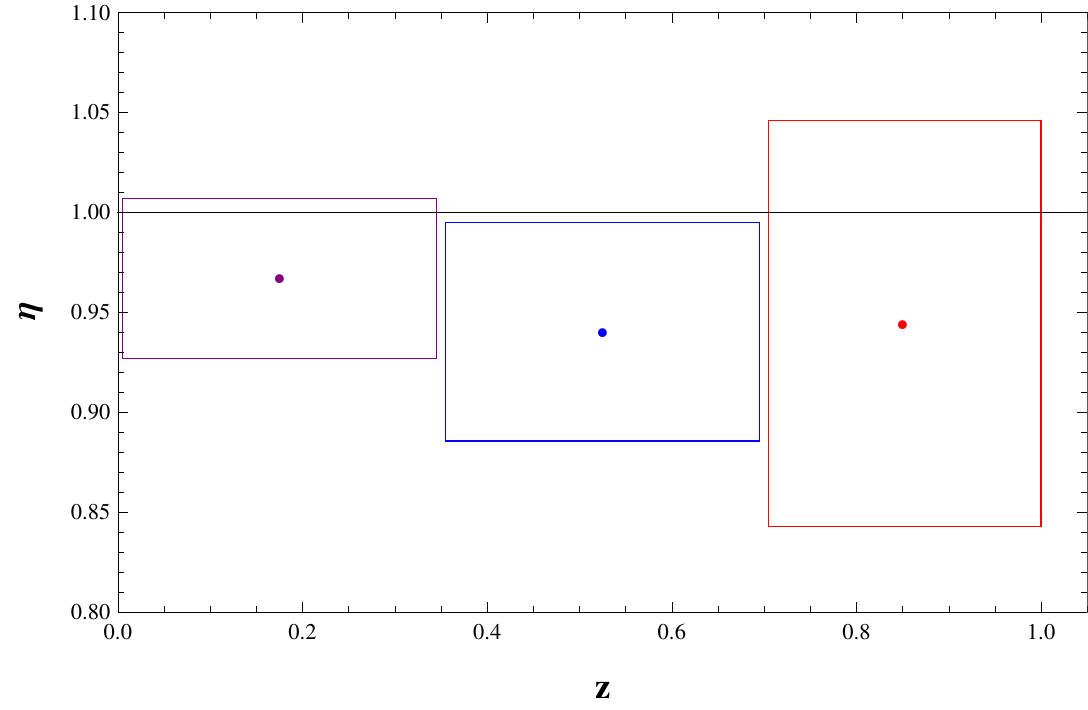}\label{fig13}}
~~~~~~~~~~~~ \subfloat[Part 3][]{\includegraphics[bb=0 0 288
286,width=2.9in]{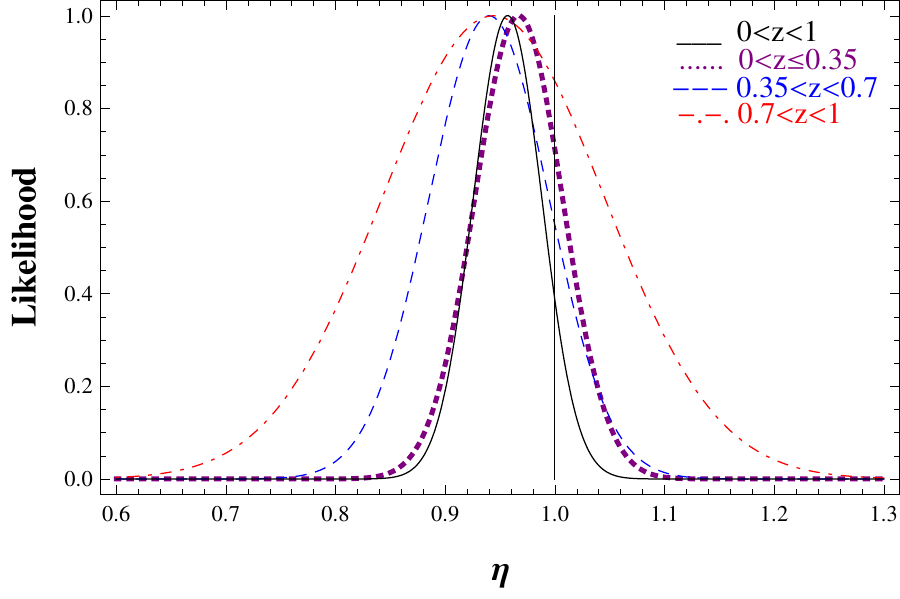}\label{fig14}} \caption{Figure (a) and (b) show
the variation of $\eta$ with z for the two parameterizations with 1$\sigma$
error bars. (c) shows piecewise constant parametrization with
1$\sigma$ error bars. (d) shows the likelihood plots for the three binned
parameters along with the constant $\eta$. All the results are for data set II.}
 \label{figa}
\end{figure}

\begin{table}[ht]
  \caption{Best fit values of $\eta_0$ with 1$\sigma$ error bars}
  \vspace{0.2cm}
  \centering
  \begin{tabular}{|c c c c|}
    \hline {\footnotesize
  $\eta (z)$} & {\footnotesize Data set $I$ } & {\footnotesize  Data set $II$ } &
  {\footnotesize  Data set $III$ } \\
    \hline

$ 1 + \eta_{0}z$  \quad & $  -0.105 \pm 0.085$ \quad&
 $ -0.098 \pm 0.084$ \quad& $ -0.066 \pm 0.070$ \quad\\

$ 1+\eta_{0} \frac{z}{1+z}$ \quad&  $  -0.161 \pm 0.134$ \quad&
 $    -0.151 \pm 0.155$ \quad&
 $  -0.103 \pm 0.106$
 \quad\\

 \hline
  \end{tabular}
  \label{tab1}
\end{table}

If we don't model $\eta$ and calculate the value of $\eta = d_L/d_A (1+z)^2$ from the data we see that a value slightly less than one is favoured. We find $\eta$ equal to $0.965 \pm 0.040$ , $ 0.971 \pm 0.037$  and  $0.966 \pm 0.070$ for the first, second and third data set respectively. This is in
agreement with earlier works in this direction. Uzan et al. \cite{uz}, found a best fit value slightly less than one and related
this trend to the systematics in the SZ/X-ray analysis of galaxy
clusters. Bassett and Kunz \cite{bk}, also found that the Supernovae
are brighter relative to the $d_{A}$ data. They suggested the gravitational lensing of high-$z$ supernovae as a possible explanation.
The constraints on the three piecewise constant $\eta_i$ for the data set II
are $\eta_1 = 0.967 \pm 0.040$ for $0 < z < 0.35$, $\eta_2 = 0.940 \pm 0.054$
for $0.35 < z < 0.7$ and $\eta_3 = 0.944 \pm 0.101$ for $0.7<z<1$. We see a decrease and then a marginal rise in $\eta$ from redshift bin one to three (figure \ref{fig13}).

Here we note that the errors on the parameters $\eta_{0}$ are
slightly improved in comparison to earlier analysis which used
SZ/X-ray measurement of galaxy cluster for $d_{A}$ \cite{h0}. The
distance duality is not accommodated within 1$\sigma$ in all three
parameterizations with the present data but the errors on the
parameters quantifying deviation from DD are still large. This is
mainly due to a few BAO data points. The constraints on the
parameters should improve with more BAO data in future observations.

\subsection{Constraints: Cosmic transparency}

The value of $\Delta \tau$ is obtained for 6 pairs of measurements
using data set II. As mentioned earlier we work within the framework of
$\Lambda$CDM
Cosmology assuming $\Omega_m$ = 0.283 $\pm$ 0.017 and $H_0$ =
69.3$\pm$ 1.5 Km/s/Mpc. The results are given in table \ref{tab3}. One can further relate the difference in opacity to
the parameter $\epsilon$ in the third parameterization considered in
the previous section as $ \tau (z) = 2 \epsilon \log(1+z)$. And if $z<<1$ then
we can approximate $\tau(z) = 2 \epsilon z$.

\begin{table}[ht]
  \caption{$\Delta \tau$ for different redshift pairs with 1$\sigma$ error bars}
  \vspace{0.2cm}
  \centering
  \begin{tabular}{|c|c|c|}
    \hline
  {Redshift} & {Pairing redshift} & {$\Delta \tau$}  \\
    \hline
0.106   \quad& 0.2 \quad& $-0.110\pm 0.193$ \quad\\
  \hline
0.2   \quad& 0.35 \quad& $-0.095\pm 0.175$ \quad\\
  \hline
0.35  \quad& 0.44 \quad& $0.180\pm 0.260$ \quad\\
  \hline
0.44  \quad& 0.57 \quad& $-0.185\pm 0.266$ \quad\\
  \hline
0.57  \quad& 0.6 \quad& $0.030\pm 0.371$  \quad\\
\hline
0.6  \quad& 0.73 \quad& $-0.015\pm 0.447$  \quad\\

 \hline
    \hline

  \end{tabular}
  \label{tab3}
   \end{table}

\begin{figure}[ht]
\centering \subfloat[Part 3][]{
\includegraphics[bb=0 0 288
286,width=3in]{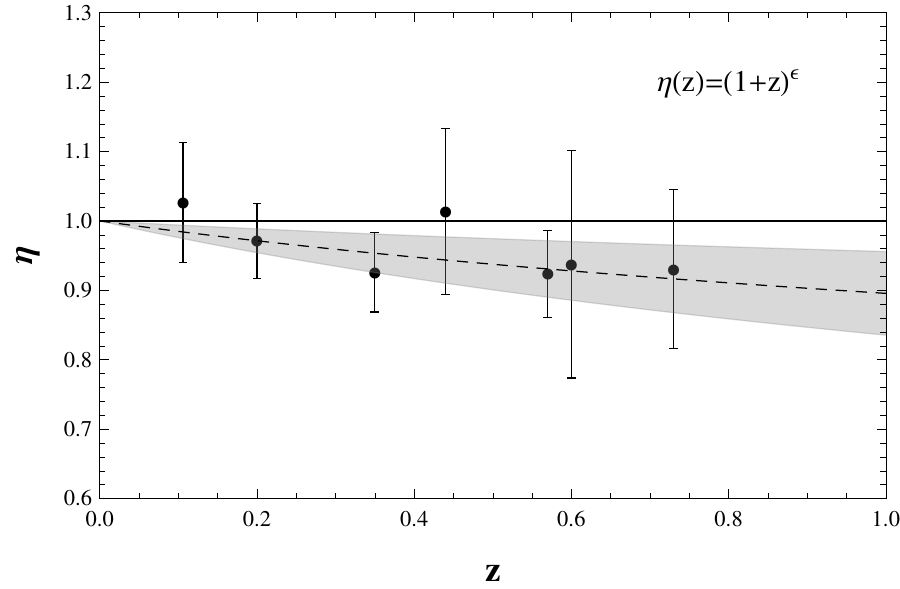}} \subfloat[Part
3][]{\includegraphics[bb=0 0 288
286,width=3in]{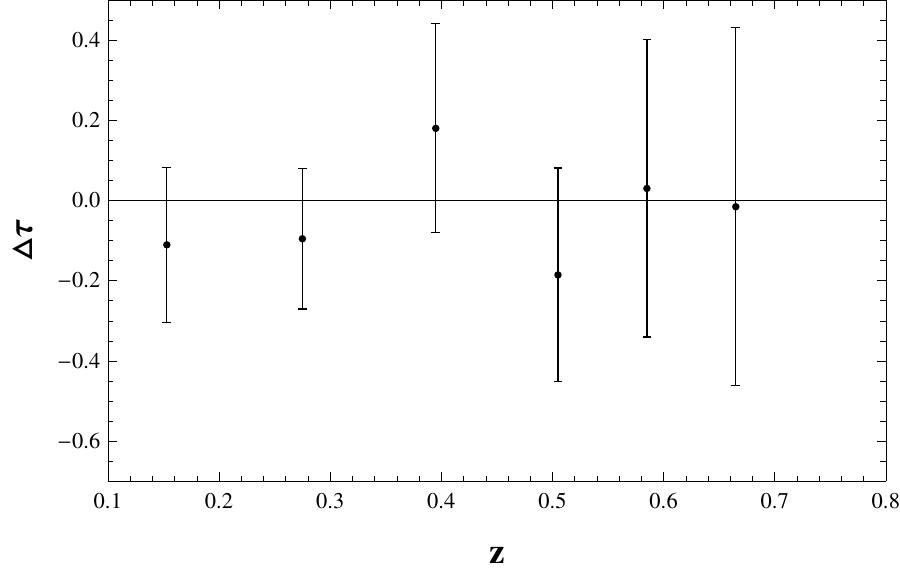}}
\caption{(a) shows the variation of $\eta$ with z for the third parameterizations with 1$\sigma$
 error bars. (b) shows the value of $\Delta\tau$ estimated in different redshift ranges with 1$\sigma$ error bars. All the results are for data set II.}
 \label{figb}
\end{figure}

The best fit value of $\epsilon$ estimated from the third parameterization is $\epsilon  = -0.159 \pm 0.097$. We see from our results that although
the errors are very large, the
best fit value of  $\Delta\tau$ is negative for a majority of
redshift pairs. Further a negative
value of $\epsilon$ is preferred by the data when the third paramterization is used. Since $\epsilon$
characterizes a departure from transparency its negative value
indicates that SNeIa are brighter than expected when estimated from
BAO data. Similar trends were reported in some of the earlier works
\cite{avg1,mor}. Figure \ref{figb} summarises the results of this section. The constraints on  $\Delta\tau$ are consistent with zero but we see that the error bars are large and there is scope for improvement from future data.

\section{Discussion}

To look for deviations from the underlying assumptions of the
Cosmology, one can device consistency checks that compare different
cosmological measurements. The distance duality relation that
derives from Etherington's relation is one such model independent
probe. In this work we have used the distance duality as a probe to
find consistency between distance measurements obtained from two
different distance indicators : SNeIa and BAO. Performing such
consistency checks can also shed light on possible systematics
present in the data. Earlier works in this direction used ADD
estimates from astrophysical sources such as galaxy clusters, FRIIb
galaxies, radio galaxies etc. Using BAO for the ADD estimate has an
advantage that it is unaffected by the astrophysical assumptions and
modeling that goes into the other probes mentioned before. Note that the distance estimates derived from BAO here are not completely model independent since we have assumed $\Lambda$CDM Cosmology to calculate $d_A$ from the $d_z$ measurement. We summarise our results below:

\begin{itemize}
\item
To perform the DD test one often applies a selection criteria to
chose the Supernovae nearest to the ADD redshift. To avoid
introducing errors due to this redshift mismatch we used the local
regression method as advocated by Cardone et al., \cite{car}. We
constrained the parameter $\eta = d_{L}/(d_{A} (1+z)^2)$ which
should be equal to 1 if DD holds and the distance measurements
agree, and found that the present data prefers a value slightly less
than 1. DD is not accommodated within 1$\sigma$, and the systematic
trend of $\eta$ being less than 1 has been reported before and needs
to be accounted for.

\item
Nesseris and Garcia-Bellido recently used Genetic Algorithm aproach to extract  model independent and bias-free reconstruction information from SNeIa, BAO and the growth rate of matter perturbations. They found a 3$\sigma$ deviation of $\eta$ from one at redshifts $z\sim 0.5$ \cite{ness}. We too observe a marginal dip in $\eta$ in the redshift bin $0.35<z<0.7$. But we require more data to derive a statistically significant conclusion.

\item
The analysis present here is
limited due to small data set. But the qualitative trend of $\eta < 1$ agrees
with earlier works in this direction. From future measurements of
BAO and SNeIa we will be able to better constrain the parameters
describing deviations from DD and hence look for violations in the
assumptions that lead to DD. We also looked
for an evidence for photon absorption by analysing the cosmic
transparency of the Universe. Assuming that there is some photon
absorption in the Universe we used the present data to constrain
its transparency. We found that the best fit to the parameter
characterizing absorption $\epsilon$, is negative indicating that
the Supernovae are brighter than expected from BAO measurements.
Various mechanism are available in the literature that could provide
a physical explanation to this: axion-photon mixing,
Chameleon-photon mixing, gravitational lensing etc.  \cite{bright}. But a probable answer
could lie in some systematic biases present in these measurements.

\end{itemize}
Tension between the SNeIa and BAO measurements have been reported
before \cite{bk,perci,la}.  We
can't constrain violation in DD or the transparency of the Universe
with much confidence with the present data. But within the framework of $\Lambda$CDM a slight disagreement
between the two different distance indicators used, is evident. We look
forward to future BAO and Supernovae data which will greatly improve
our constraints on these parameters.

\section*{Acknowledgment}

It is a pleasure to thank Bruce Bassett, Chris Blake and Martin Kunz for discussion and suggestions. One of the author (D.J.) thanks A. Mukherjee and S. Mahajan for
providing the facilities to carry out the research. R.N.
acknowledges support under CSIR - SRF scheme (Govt. of India).
Authors also acknowledge the financial support provided by
Department of Science and Technology, Govt. of India under project
No. SR/S2/HEP-002/008.


\begin{thebibliography}{99}

\bibitem{sur}
Sloan Digital Sky Survey (SDSS): www.sdss3.org;

Giant Magellan Telescope (GMT): www.gmto.org;

James Webb Space Telescope (JWST): www.jwst.nasa.gov;

Euclid survey: http://sci.esa.int/euclid;

Dark Energy Survey: www.darkenergysurvey.org;

Large Synoptic Survey Telescope (LSST): www.lsst.org

Planck: www.rssd.esa.int/planck

\bibitem{DHwein} Weinberg D. H. et al., {\it Observational Probes of Cosmic Acceleration} [arXiv:1201.2434].

\bibitem{exp}

Riess A. G. et al., (Supernova Search Team) {\it Observational
evidence from Supernovae for an accelerating Universe \& a
cosmological constant}, \apj~ {\bf 116}, (1998) 1009,
[arXiv:astro-ph/9805201];

Perlmutter S. et al., (Supernova Cosmology Project) {\it Measurement
of Omega and Lambda from 42 high-redshift Supernovae}, \apj~ {\bf
517}, (1999) 565, [arXiv:astro-ph/9812133];

Astier P. et al., {\it The Supernovae Legacy Survey: measurement of
$\Omega_m$, $\Omega_\lambda$ and $w$ from the first year data set},
\AnA ~ {\bf 447}, (2006) 31, [arXiv:astro-ph/0510447];

Padmanabhan T., {\it Cosmological constant: The weight of the
vacuum, Phys. Rep.}, {\bf 380} (2003) 235 [arXiv:hep-th/0212290]
[SPIRES];

Sahni V. \& Starobinsky A., {\it Reconstructing Dark Energy},
\ijmpd~ {\bf 15} (2006) 2105 [arXiv:astro-ph/0610026] [SPIRES];

Frieman J. A., Turner M. \& Huterer D., {\it Dark Energy and the
Accelerating Universe}, \ARAnA~ {\bf 46} (2008) 385
[arXiv:0803.0982] [SPIRES];

Caldwell R. R. \& Kamionkowski M., {\it The Physics of Cosmic
Acceleration, Ann. Rev. Nucl. Part. Sci.}, {\bf 59} (2009) 397
[arXiv:0903.0866] [SPIRES].

\bibitem{ruth} Durrer R. {\it What do we really know about Dark Energy?}, {\it Philosophical Transactions of the Royal Society A: Mathematical, Physical and Engineering Sciences} {\bf 369}
(2011) 5102 [arXiv:1103.5331];

Bianchi E. \& Rovelli C., {\it Why all these prejudices against a
constant?} [arxiv:1002.3966].

\bibitem{wein} Weinberg S., {\it The cosmological constant problem},
{\it Rev. Mod. Phys.}, {\bf 61} (1989) 1.

\bibitem{sami} Copeland E.J. et al., {\it Dynamics of dark energy},
\ijmpd~ {\bf D15} (2006) 1753.

\bibitem{kunz} Kunz M., {\it The phenomenological approach to
modeling the dark energy} [arXiv:1204.5482].

\bibitem{good} Goodman G., {\it Geocentrism reexamined},
\prd~ {\bf 52} (1995) 1821.

\bibitem{cald} Caldwell R.R. \& Stebbins A., {\it A Test of
the Copernican Principle}, \prl~ {\bf 100} (2008) 191302.

\bibitem{clark} Clarkson C., Bassett B \& Lu T.H.C.,
{\it A general test of the Copernican Principle}, \prd~ {\bf 101}
(2008) 011301.

\bibitem{sri} Srianand R. et al., {\it The cosmic microwave
background radiation temperature at a redshift of 2.34
}, \nature~ {\bf 408} (2000) 435.

\bibitem{sandage} Lubin L.M \& Sandage A., {\it The Tolman
Surface Brightness Test for the Reality of the Expansion. IV. A
Measurement of the Tolman Signal and the  Luminosity Evolution of
Early-Type Galaxies }, {\it AJ}, {\bf 122} (2001) 1084.


\bibitem{mor} More S., Bovy J. \& Hogg D.W.,  {\it Cosmic
transparency: A test with the baryon acoustic feature and type- Ia
supernovae}, \apj~ {\bf 696} (2009) 1727 [arXiv:0810.5553] [SPIRES].


\bibitem{avg1} Avgoustidis A., Verde L. \& Jimenez R.,
{\it Consistency among distance measurements:transparency, BAO scale
and accelerated expansion}, JCAP~  {\bf 06} (2009) 012
[arXiv:0902.2006] [SPIRES].


\bibitem{bk} Bassett B. A. \& Kunz M., {\it Cosmic
distance-duality as a probe of exotic physics and acceleration},
\prd~ {\bf 69} (2004) 101305.

\bibitem{axion}

Bassett B. \& Kunz M., {\it Cosmic Acceleration versus Axion-Photon Mixing
}, \apj~ {\bf 607} (2004) 661;

Mirizzil A. et al.,
{\it Photon-axion conversion as a mechanism for supernova dimming:
Limits from CMB spectral distortion}, \prd~ {\bf 72} (2005) 023501;

Avgoustidis A. et al., {\it Constraints on cosmic opacity and
beyond the standard model physics from cosmological distance
measurements}, JCAP~ {\bf 10} (2010) 024.

\bibitem{eth} Etherington I. M. H., {\it Phil. Mag.} {\bf 15}
(1933) 761 (Reprinted in : \grg {\bf 39} (2007) 1055).


\bibitem{uz} Uzan J. P. et al., {\it The distance duality
relation from X-ray and SZ observations of clusters}, \prd~
{\bf 70} (2004) 083533 [astro-ph/0405620].

\bibitem{dd} Temple G., {\it New Systems of Normal Co-ordinates for Relativistic Optics, Proc. Roy.
Soc. Ser.}  A {\bf 168} (1938) 122;

Kristian J. \& Sachs R. K., {\it Observations in Cosmology}, \apj~ {\bf 143} (1966) 379 [SPIRES];

Linder E. V., {\it Isotropy of the microwave background by gravitational lensing}, \AnA~ {\bf 206} (1988) 190;

Schneider P., Ehlers J. \& Falco E. E., {\it Gravitational Lenses} Springer-Verlag, Berlin Germany (1992).

\bibitem{ellisgrg} Ellis G. F. R., {\it On the definition of distance in general relativity: I.M.H. Etherington (Philosophical Magazine ser. 7, vol. 15, 761 (1933))}, \grg~ {\bf 39} (2007) 1047.

\bibitem{b} Bernardis F. De, Giusarma E. \& Melchiorri A., {\it Constraints on distance duality relation from Sunyaev Zeldovich effect and Chandra X-ray measurements}, \ijmpd~ {\bf 15} (2006) 759 [arXiv:gr-qc/0606029] [SPIRES].

\bibitem{co} Corasaniti P. S., {\it The Impact of Cosmic Dust on Supernova Cosmology}, \mnras~ {\bf 372} (2006) 191 [arXiv:astro-ph/0603883] [SPIRES].

\bibitem{la} Lazkoz R., Nesseris S. \& Perivolaropoulos L., {\it Comparison of Standard Ruler and Standard Candle constraints on Dark Energy Models}, JCAP~ {\bf 07} (2008) 012 [arXiv:0712.1232] [SPIRES].

\bibitem{h0} Holanda R. F. L., Lima J.A.S. \& Ribeiro M.B., {\it Testing the Distance-Duality Relation with Galaxy Clusters and Type Ia Supernovae}, \apj~ {\bf 722} (2010) L233 [arXiv:1005.4458] [SPIRES];


\bibitem{us} Nair R., Jhingan S., Jain D., {\it  Observational cosmology and the cosmic distance duality relation}, JCAP~ {\bf 05} (2011) 023 [arXiv:1102.1065].

\bibitem{restdd}
Holanda R. F. L., Lima J.A.S. \& Ribeiro M.B., {\it Cosmic distance duality relation and the shape of galaxy clusters}, \AnA~ {\bf 528}  (2011) L14;

Li Z. et al., {\it  Cosmological-model-independent tests for the distance-duality relation from Galaxy Clusters and Type Ia Supernova}, \apj~ {\bf 729} (2011) L14;

Liang N. et al., {\it A Consistent Test of the Distance-Duality Relation with Galaxy Clusters and Type Ia Supernave} [arXiv: 1104.2497];

Meng X.L. et al., {\it Morphology of Galaxy Clusters: A Cosmological Model-Independent Test of the Cosmic Distance-Duality Relation}, \apj~ {\bf 745} (2012) 98 [arXiv: 1104.2833];

Cao S. \& Liang N. {\it Testing the Distance-Duality Relation with a Combination of Cosmological Distance Observations}, {\it Research in Astronomy and Astrophysics} {\bf 11} (2011) 1199 [arXiv:1104.4942];

Lima J.A.S. et al., {\it Deformed Distance Duality Relations and Supernova Dimming},
 \apjl ~ {\bf 742} (2011) L26;

Goncalves R. S., Holanda R. F. L. \& Alcaniz J.S, {\it Testing the cosmic distance duality with X-ray gas mass fraction and supernovae data} \mnras~ {\bf 420} (2012) L43;

Holanda R. F. L., Goncalves R. S., \& Alcaniz J.S, {\it A test for cosmic distance duality} JCAP~ {\bf 06} (2012) 022;

Holanda R. F. L., {\it Constraints on the Hubble Parameter from galaxy clusters and the Validity of the Cosmic Distance Duality Relation} \ijmpd ~ {\bf 21} (2012) [arXiv:1202.2309].

\bibitem{koma}
Page L. et al., {\it First-Year Wilkinson Microwave Anisotropy Probe (WMAP) Observations: Interpretation of the TT and TE Angular Power Spectrum Peaks}, \apjs~ {\bf 148} (2003) 233;

Spergel D.N., {\it Three-Year Wilkinson Microwave Anisotropy Probe (WMAP) Observations: Implications for Cosmology}, \apjs~ {\bf 170} (2007) 377;

Komatsu E. et al., {\it Five-Year Wilkinson Microwave Anisotropy Probe Observations:
Cosmological Interpretation}, \apj~ {\bf 180} (2009) 330, [arXiv:0803.0547].

\bibitem{mather} Mather J.C. et al., {\it Measurement of the cosmic microwave background spectrum by the COBE FIRAS instrument} \apj~ {\bf 420} (1994) 439.


\bibitem{avg2} Avgoustidis A. et al., {\it Constraints on the CMB temperature-redshift dependence from SZ and distance measurements}, JCAP {\bf 02} (2012) 013.

\bibitem{bl1} Blake C. et al., {\it The WiggleZ Dark Energy Survey: mapping the distance-redshift relation with
baryon acoustic oscillations}, \mnras ~ {\bf 418} (2011) 1707 [arXiv:1108.2635].

\bibitem{bl2} Blake C. et al., {\it The WiggleZ Dark Energy Survey: testing the cosmological model with baryon
acoustic oscillations at z=0.6}, \mnras ~ {\bf 415} (2011) 2892 [arXiv:1105.2862].

\bibitem{eis} Eisenstein et al., {\it Cosmic Complementarity: $H_{o}$ and $\Omega_{m}$ from Combining CMB Experiments
and Redshift Surveys}, \apj~ {\bf 504} (1998) 57, [arXiv:astro-ph/9805239].

\bibitem{coo} Cooray A. et al., {\it Measuring Angular Diameter Distances through Halo Clustering} \apjl~ {\bf 557} (2001) 7, [arXiv:astro-ph/0105061].

\bibitem{car} Cardone V.F. et al., {\it Testing the distance duality relation with present and future data} [arXiv:1205.1908].

\bibitem{aman} Amanullah R. et al., {\it Spectra and Light Curves of Six Type Ia Supernovae at $0.511< z <1.12$
and the Union2 Compilation}, \apj~ {\bf 716} (2010) 712, [arXiv:1004.1711].


\bibitem{perc} Percival W. J. et al., {\it Baryon acoustic oscillations in the Sloan Digital Sky Survey Data
Release 7 galaxy sample}, \mnras~ {\bf 401} (2010) 2148, [arXiv:0907.1660].

\bibitem{beut} Beutler F. et al., {\it The 6dF Galaxy Survey: baryon acoustic oscillations and the local Hubble
constant}, \mnras~ {\bf 416} (2011) 3017, [arXiv:1106.3366].

\bibitem{boss1} Anderson L. et al., {\it The clustering of galaxies in the SDSS-III Baryon Oscillation
Spectroscopic Survey: Baryon Acoustic Oscillations in the Data
Release 9 Spectroscopic Galaxy Sample}, [arXiv:1203.6594].

\bibitem{boss} Reid B. A. et al., {\it The clustering of galaxies in the SDSS-III Baryon Oscillation
Spectroscopic Survey: measurements of the growth of structure and
expansion rate at $z=0.57$ from anisotropic clustering
}, [arXiv:1203.6641].

\bibitem{xu} Xu X. et al., {\it Measuring $D_A$ and $H$ at $z = 0.35$ from the SDSS DR7 LRGs using baryon acoustic oscillations}, [arXiv: 1206.6732].

\bibitem{bl3} Blake C. et al., {\it The WiggleZ Dark Energy Survey: Joint measurements of the expansion and growth history at $z < 1$}, [arXiv: 1204.3674].

\bibitem{ness} Nesseris S. \& Garcia-Bellido J. {\it A new perspective on Dark Energy modeling via Genetic Algorithms}, [arXiv: 1205.0364]

\bibitem{bright} Burrage C., {\it Supernova brightening from chameleon-photon mixing} \prd~ {\bf 77} (2008) 043009;

Jaeckel J. \& Ringwald A., {\it The Low-Energy Frontier of Particle Physics, Ann. Rev. Nucl. Part. Sci} {\bf 60} (2010) 405 [arXiv:1002.0329].

\bibitem{perci} Percival W. J. et al., {\it Measuring the Baryon Acoustic Oscillation scale using the Sloan Digital Sky Survey and 2dF Galaxy Redshift Survey}, \mnras~ {\bf 381} 1053 (2007).



\end{thebibliography}
\end{document}